\documentclass[10pt,preprint]{aastex}

\begin{document}

\shortauthors{Luhman}
\shorttitle{Wide Binary Brown Dwarf}

\title{The First Discovery of a Wide Binary Brown Dwarf\altaffilmark{1}}

\author{K. L. Luhman}
\affil{Harvard-Smithsonian Center for Astrophysics, 60 Garden Street,
Cambridge, MA 02138}

\email{kluhman@cfa.harvard.edu}

\altaffiltext{1}
{Based on observations performed at Las Campanas Observatory.
This publication makes use of data products from the Two Micron All
Sky Survey, which is a joint project of the University of Massachusetts
and the Infrared Processing and Analysis Center/California Institute
of Technology, funded by the National Aeronautics and Space
Administration and the National Science Foundation.}

\begin{abstract}

I present observations of a new faint double, 2MASS~J11011926-7732383AB,
toward the Chamaeleon~I star-forming region.
From optical and near-infrared images of the pair, I measure a separation
of $1\farcs44$ and extract $RIJHK_s$ photometry of the components 
($I_A=17.21$, $\Delta I=1.07$, $K_{s A}=11.97$, $\Delta K_s=0.84$).
I use resolved optical spectroscopy to derive spectral types of M7.25 and M8.25
for the A and B components, respectively.
Based on the strengths of gravity-sensitive features in these data, such as 
the Na~I and K~I absorption lines, 
I conclude that these objects are young members of
Chamaeleon~I rather than field stars.
The probability that this pair is composed of unrelated late-type 
members of Chamaeleon~I is low enough ($\sim5\times10^{-5}$) to definitively 
establish it as a binary system.
After estimating extinctions, effective temperatures, and bolometric 
luminosities for the binary components, I place them on the H-R 
diagram and infer their masses with the evolutionary models of Chabrier and 
Baraffe, arriving at substellar values of 0.05 and 0.025~$M_{\odot}$.
The projected angular separation of this system corresponds to 240~AU
at the distance of Chamaeleon~I, making it the first known
binary brown dwarf with a separation greater than 20~AU. 
This demonstration that brown dwarfs can form in fragile, easily 
disrupted configurations is direct evidence that the formation of brown dwarfs 
does not require ejection from multiple systems or other dynamical effects.
It remains possible that ejection plays a role in the formation of 
some brown dwarfs, but it is not an essential component according to these
observations.

\end{abstract}


\section{Introduction}
\label{sec:intro}

Over the past decade, hundreds of free-floating brown dwarfs have been 
discovered in the field and in young clusters \citep{bas00}, but 
their origin remains a mystery.
In recent numerical hydrodynamical simulations of the fragmentation of 
molecular cloud cores, self-gravitating objects can form with 
initial masses of only $\sim1$~$M_{\rm Jup}$, but these fragments continue to
accrete matter from their surrounding core, usually to the point of eventually
reaching stellar masses \citep{bos01,bat03}. As a result, the formation 
of brown dwarfs appears difficult to achieve in these models. One possible 
solution has been proposed by \citet{rc01} and \citet{bos01}, and 
investigated further by \citet{bat02} and \citet{kb03b}.
In this scenario, dynamical interactions within a small cluster of protostellar
sources leads to the ejection of one of the embryos, which 
prematurely halts accretion from the core's reservoir of infalling material 
and results in the formation of a brown dwarf.

Models for the origin of brown dwarfs can be tested 
through measurements of the multiplicity of substellar objects. 
Significant progress in this area has been made through extensive
direct imaging of primaries in the field with masses near and below the 
hydrogen burning mass limit
\citep{koe99,mar99,rei01b,clo02a,clo02b,clo03,bou03,bur03,fre03,giz03}.
In the data from these surveys, binary brown dwarfs occur at a rate of 
$\sim15$\% for separations of $a>1$~AU and exhibit
maximum separations of $a\sim20$~AU. 
Ejection models predict a somewhat lower binary fraction ($\sim5$\%) but
a similar maximum separation ($a\sim10$~AU) \citep{bat02}.
\citet{bur03} concluded that the observed maximum separation is not a 
reflection of disruption of wider binaries by interactions with stars 
or molecular clouds, which is supported by the absence of wide binaries 
among substellar primaries in less evolved populations in open clusters
\citep{mar98,mar00,mar03} and in star-forming regions \citep{neu02,bou04}.
Instead, \citet{bur03} suggested that wide low-mass binaries do not form or 
are disrupted at a very early stage.

I report the discovery of the first known widely-separated 
binary brown dwarf, which was found serendipitously during observations of
candidate young brown dwarfs in the Chamaeleon~I star-forming region. 
I present optical and near-IR images and optical spectroscopy of the
components of the pair (\S~\ref{sec:obs}), assess the
membership of the two objects in Chamaeleon~I (\S~\ref{sec:mem}) 
and in a binary system (\S~\ref{sec:bin}), place them on the 
Hertzsprung-Russell (H-R) diagram (\S~\ref{sec:ext}), estimate their masses from
theoretical evolutionary models (\S~\ref{sec:mass}), and discuss the 
implications of this new binary system for theories of brown dwarf formation
(\S~\ref{sec:dis}).

\section{Observations}
\label{sec:obs}

During spectroscopy of candidate young brown dwarfs in Chamaeleon~I on
2004 January 10 with the Inamori Magellan Areal Camera and Spectrograph 
(IMACS) on the Magellan~I telescope at Las Campanas Observatory,
I found that one of the candidates, 2MASS~J11011926-7732383 (hereafter
2M~1101-7732), appeared as a 
$\sim1\arcsec$ double on the telescope acquisition camera. I obtained an 
optical spectrum of the brighter component, found it to be indicative of a 
young late-type object, and thus proceeded to perform the following 
observations to study the pair in more detail.

\subsection{Photometry}
\label{sec:phot}

Optical images of 2M~1101-7732AB were obtained with IMACS 
on the Magellan~I telescope on the night of 2004 January 11.
The instrument contained eight $2048\times4096$ CCDs separated by
$\sim10\arcsec$ and arranged in a $4\times2$ mosaic.
The plate scale was $0\farcs202$~pixel$^{-1}$.
One and two 60~sec exposures were acquired of the target through $I$ and $R$ 
filters, respectively. 
I obtained near-infrared (IR) images of 2M~1101-7732AB with
Persson's Auxiliary Nasmyth Infrared Camera (PANIC) on the Magellan~I telescope 
on the night of 2004 April 7. The instrument contained one $1024\times1024$ 
HgCdTe Hawaii array with a plate scale of $0\farcs126$~pixel$^{-1}$.
For each of the filters $J$, $H$, and $K_s$,
one exposure of the target was taken at each position in a $3\times3$
dither pattern with dither sizes of $5\arcsec$. 
The individual exposure times were 3, 4, and 4~sec for the three filters,
respectively.

\begin{figure}
\plotone{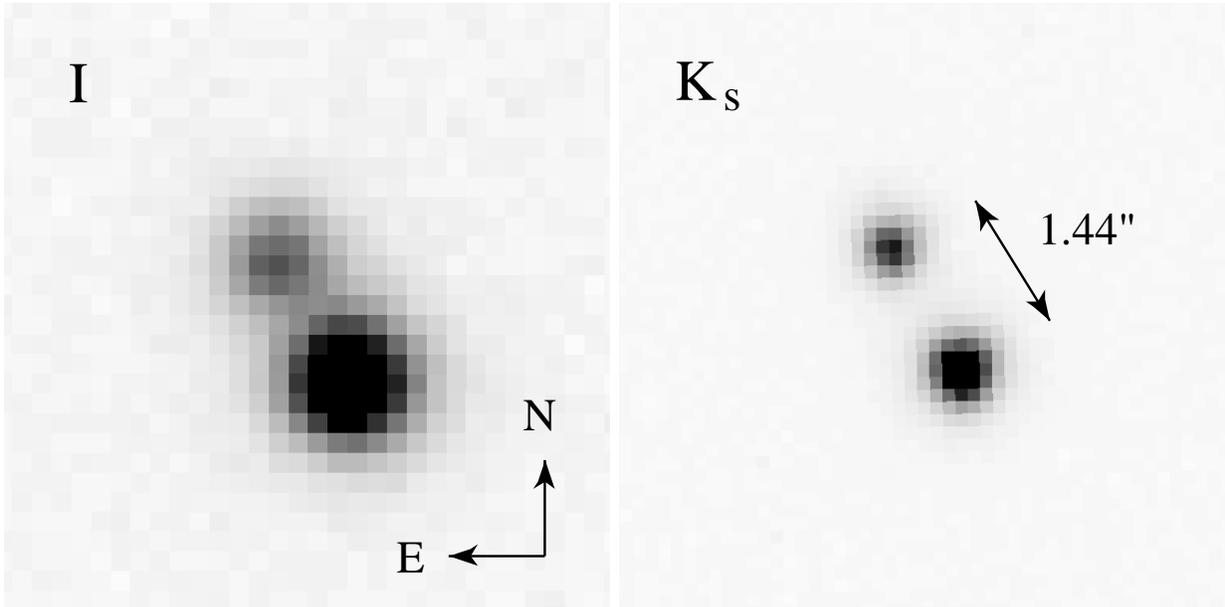}
\caption{
Images of the binary system 2M~1101-7732AB at $I$ and $K_s$
(FWHM$=0\farcs85$ and $0\farcs39$). 
Each image is $3\arcsec$ on a side and is displayed linearly from the 
background level to half of the peak flux of the primary.
}
\label{fig:image}
\end{figure}

After the data at each filter were dark subtracted, flat fielded, and 
combined, the final optical and IR images exhibited FWHM$=0\farcs85$ and 
$0\farcs45$ for point sources, respectively.
I extracted photometry of the two components of 2M~1101-7732AB with
the task PHOT under the IRAF package APPHOT using radii of three and four 
pixels in the optical and IR images.
The optical photometry was calibrated in the Cousins $I$ system by combining
data for standards across a range of colors \citep{lan92} with the
appropriate aperture and airmass corrections.
The IR photometry was calibrated with measurements from the 2MASS Point Source
Catalog for the nine 2MASS sources surrounding 2M~1101-7732AB in these images.
Plate solutions were derived from coordinates measured in the 2MASS Point 
Source Catalog for stars that appeared in the images and were not saturated
(excluding the 2M~1101-7732AB, which is unresolved in 2MASS data).
Because the optical images encompass
a larger area and thus contain more 2MASS sources than the IR data, 
the absolute astrometry of 2M~1101-7732AB is better determined in
the optical images. Meanwhile, the higher resolution of the IR data provides
greater precision in measurements of the relative coordinates of A and B.
Therefore, the astrometry of A was measured from the $I$-band image and 
the offset of B from A was taken as the average value measured in the 
$J$, $H$, and $K_s$ images. 
The absolute and relative uncertainties in the coordinates are $\pm0\farcs1$
and $\pm0\farcs02$, respectively.
The astrometric and photometric measurements of 2M~1101-7732AB
are listed in Table~\ref{tab:data}. 
A second set of images at $K_s$ (FWHM=$0\farcs39$) were obtained on 2004 
April 9, which produced photometry at $K_s$ and relative coordinates of 
A and B that agreed with those in Table~\ref{tab:data} to within 0.02~mag 
and $0\farcs01$. 
The magnitudes measured here for A+B are brighter than those from the 2MASS 
Point Source Catalog by 0.1, 0.05, and 0.07~mag at $J$, $H$, and $K_s$, 
respectively. 
The images of 2M~1101-7732AB at $I$ and $K_s$ are shown 
in Figure~\ref{fig:image}.

\subsection{Spectroscopy}
\label{sec:spec}

\begin{figure}
\epsscale{0.5}
\plotone{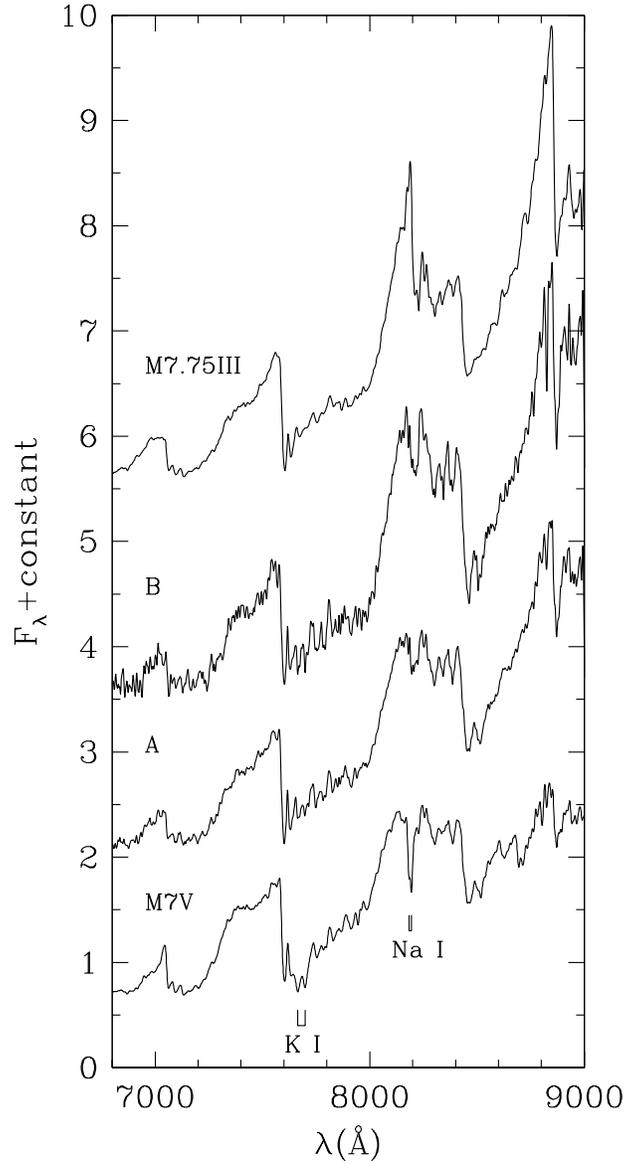}
\caption{
Spectra of 2M~1101-7732 A and B (M7.25 and M8.25) compared to 
data for the field dwarf vB~8 (M7V) and the field giant YY~Cha (M7.75III).
The strengths of gravity-sensitive features such as Na~I and K~I in the
spectra of 2M~1101-7732 A and B are intermediate between those
of the dwarf and the giant, indicating that A and B are pre-main-sequence 
objects and thus are members of the Chamaeleon~I star-forming region. 
The spectra are displayed at a resolution of 10~\AA\ and are normalized at
7500~\AA.}
\label{fig:spec}
\end{figure}

I performed optical spectroscopy on 2M~1101-7732AB
with IMACS on the Magellan~I telescope on the night of 2004 April 8. 
The spectrograph was operated with the 200~l~mm$^{-1}$ grism, OG570
blocking filter, and $1\farcs0$ long slit (FWHM$=8$~\AA).
One 30~min exposure was obtained of 2M~1101-7732AB with the slit
placed along the axis connecting the components, corresponding to a position
angle of $30\arcdeg$. 
In the resulting images, the components exhibited FWHM$=1\arcsec$, and thus
were sufficiently resolved from each other for the extraction of separate
spectra. On a coordinate system where the primary is at the origin and the
secondary is at $+1.44\arcsec$, the spectra of the primary and secondary 
were extracted from apertures of $-1\arcsec$ to $+0\farcs6$ and $+0\farcs8$ to
$+2\farcs4$, respectively. To correct for contamination by the primary in 
the latter aperture, the same aperture on the opposite side of the primary
($-2\farcs4$ to $-0\farcs8$) was used for background subtraction.
Because the slit was not aligned at the parallactic angle, these data are
subject to differential slit loss with wavelength. To correct for this effect,
the spectra were multiplied by a spectral function such that the primary's
spectrum matched the data for the primary from 2004 January 10, which was
acquired with the slit at the parallactic angle. 
The spectra of 2M~1101-7732 A and B are shown in 
Figure~\ref{fig:spec} after smoothing to a resolution of 10~\AA. 
For comparison, I include spectra of the field dwarf vB~8 and the field giant
YY~Cha. The first spectrum was obtained with the same 
instrument configuration (except with a $0\farcs9$ slit) on 2004 April 27 
and the second spectrum was measured by \citet{luh04a}.

\section{Analysis}

\subsection{Evidence of Membership in Chamaeleon~I}
\label{sec:mem}

\begin{figure}
\plotone{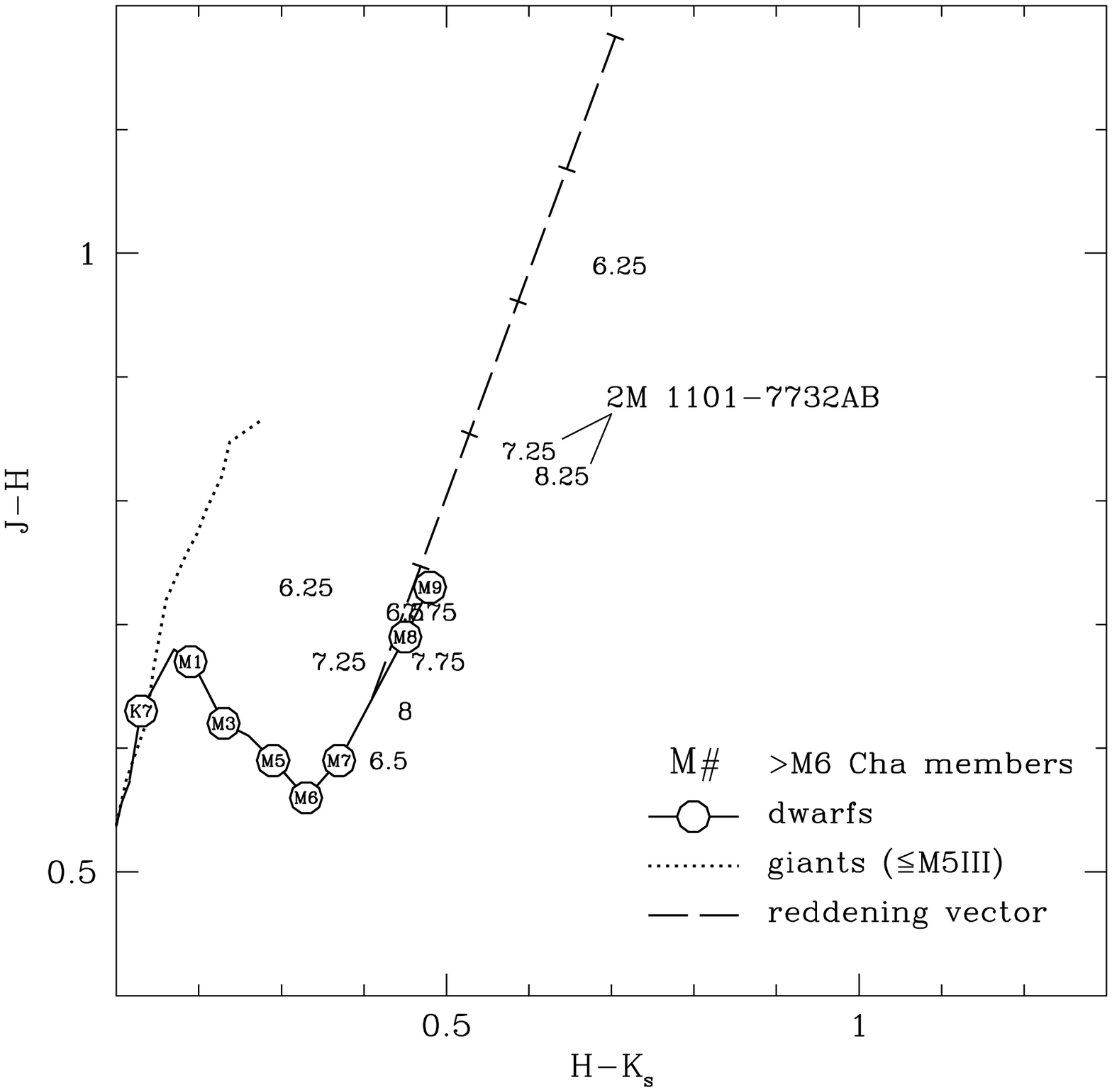}
\caption{
$H-K_s$ versus $J-H$ for 2M~1101-7732 A and B and the eight known 
members of Chamaeleon~I with spectral types later than M6 from \citet{luh04a}.
These sources are represented by the M subclass of their spectral types.
I include sequences for typical field dwarfs ({\it solid line}; $\leq$M9V, 
\citet{leg92}) and giants ({\it dotted line}; $\leq$M5~III, \citet{bb88}) 
and a reddening vector
originating at M7.5V with marks at intervals of $A_V=1$ ({\it dashed line}).
The separation of 2M~1101-7732 A and B from the dwarf sequence
is indicative of some combination of reddening, excess emission from
circumstellar material, and departure of the intrinsic colors from dwarf
values, each of which is evidence of the youth and thus the membership in 
Chamaeleon~I of these sources.}
\label{fig:jhhk}
\end{figure}

I now use the photometry and spectroscopy from the previous section to 
assess membership in Chamaeleon~I and measure spectral types for 
the components of 2M~1101-7732AB by applying the methods 
described in my previous studies of young populations 
\citep{luh99,luh04a,luh03a,luh03b}.
The spectra of 2M~1101-7732 A and B exhibit evidence of
youth, and thus membership in Chamaeleon~I, in the form of K~I and 
Na~I line strengths that are intermediate between those of field dwarfs
and giants \citep{mar96,luh99}, as illustrated in Figure~\ref{fig:spec} by
the comparison to vB~8 (M8V) and YY~Cha (M7.75III, \citet{luh04a}).
Other features in these spectra are also distinctive from field objects
and clearly indicative of pre-main-sequence surface gravities, such 
as the CaH band at 7000~\AA\ and the shape of the spectrum across 8200~\AA.
Indeed, the spectra for 2M~1101-7732 A and B agree well with those of 
other known late-type members of star-forming regions. 
In the diagram of $H-K_s$ versus $J-H$ in Figure~\ref{fig:jhhk}, the colors
of 2M~1101-7732 A and B differ significantly from those of field dwarfs, which
indicates the presence of reddening or circumstellar material, or a departure
in the intrinsic colors from dwarf values, any one of which is further
evidence of membership in Chamaeleon~I. 
Through a comparison to averages of dwarfs and giants \citep{luh99}
and to previously observed young late-type objects \citep{bri02,luh04a}, 
I measure spectral types of M7.25 and M8.25 for 2M~1101-7732 A and B.

\subsection{Evidence of Binarity}
\label{sec:bin}

\begin{figure}
\epsscale{1}
\plotone{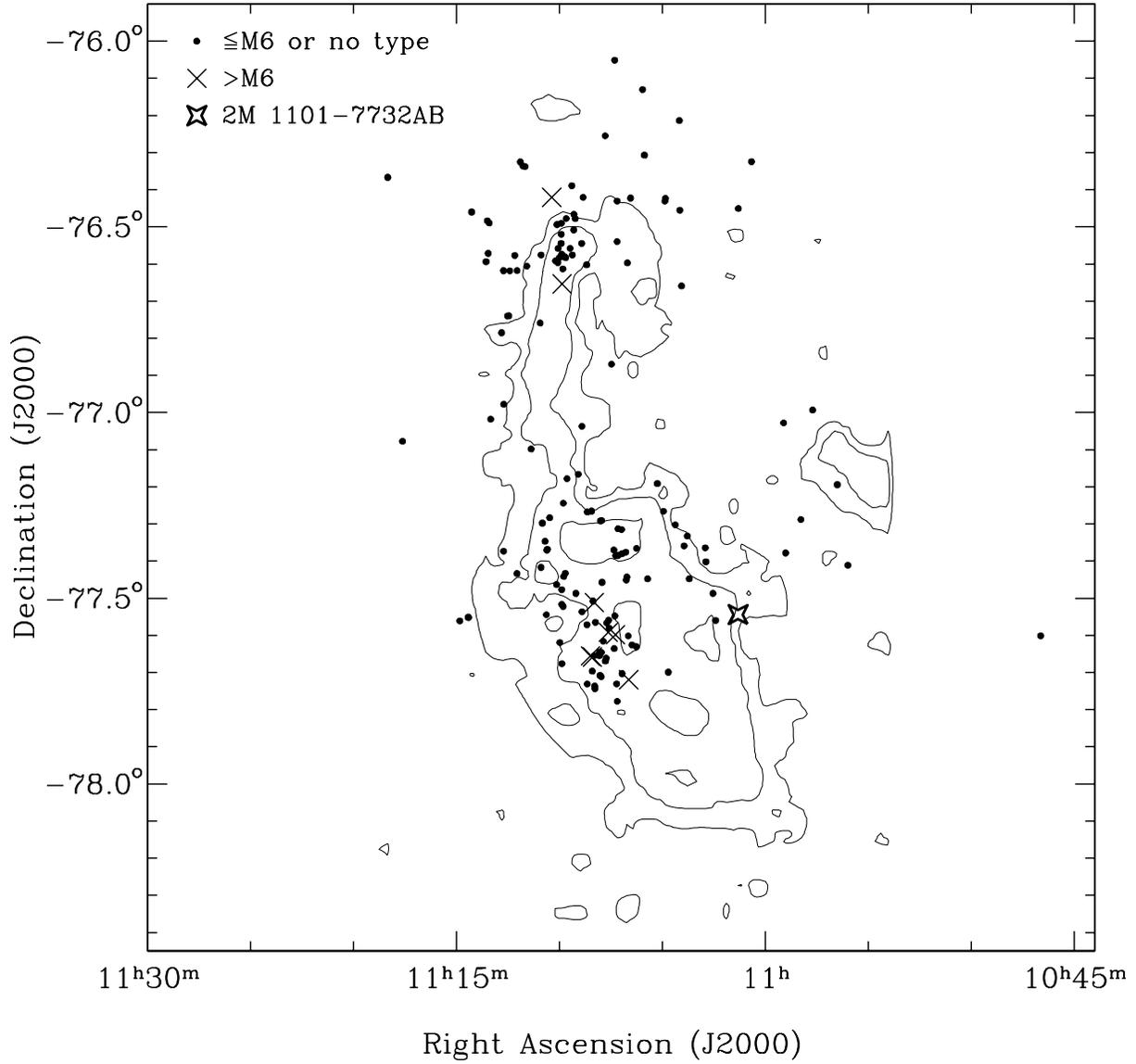}
\caption{Position of 2M~1101-7732AB relative to the known
members of the Chamaeleon~I star-forming region \citep{luh04a}.
The contours represent the extinction map of \citet{cam97} at intervals
of $A_J=0.5$, 1, and 2.}
\label{fig:map}
\end{figure}

Because of the low stellar density of the Chamaeleon~I star-forming region,
it is unlikely that unrelated members will be seen in projection near each
other. This applies particularly to the vicinity of 2M~1101-7732AB, which is
a sparsely populated area of the cloud, as shown in Figure~\ref{fig:map}.
The nearest known member to 2M~1101-7732AB is at a distance of $4\arcmin$.
According to the results of the survey in which 2M~1101-7732AB was discovered,
the area of 2~deg$^{2}$ encompassing Chamaeleon~I contains 
$\sim20$ young brown dwarfs down to the mass of 2M~1101-7732B
\footnote{Figure~\ref{fig:map} includes only the members compiled by
\citet{luh04a} and not the brown dwarfs found in the new survey.}.
The probability that two of these brown dwarfs have a projected separation 
of $a\leq1\farcs44$ is $5\times10^{-5}$. Based on this low probability,
I conclude that 2M~1101-7732 A and B comprise a bound binary system rather than
two unrelated members of Chamaeleon~I.
To test the binarity of this pair through common proper motions, such
measurements would need to have extremely high precision since unrelated 
members of the cluster share the same motion to within $\sim2$~km~s$^{-1}$, or 
$\sim0\farcs0025$~yr$^{-1}$.

\subsection{Extinctions, Temperatures, and Luminosities}
\label{sec:ext}

Extinctions for 2M~1101-7732 A and B have been
estimated from the spectra in the manner described by \citet{luh04a}.
The resulting values of $A_J=0.45$ and 0 ($\lesssim0.2$) are consistent with 
the extinctions implied by the color excesses of $E(R-I)=0.2$ and 0.1 produced 
when the observed colors are combined with estimates of intrinsic values of 
$R-I$ for young late-type objects \citep{luh03b}. 
When the observed near-IR colors of A are dereddened by $A_J=0.45$, 
there remain small excesses of $E(J-H)=0.06$ and $E(H-K_s)=0.11$ relative 
to dwarf values.
The IR colors of B exhibit larger excesses of $E(J-H)=0.12$ and $E(H-K_s)=0.18$.
These excesses could result from circumstellar material or from 
deviations of the intrinsic colors from the average dwarf values. 
The fact that these excesses remain when computed relative to other late-type 
members of Chamaeleon~I instead of field dwarfs, as illustrated
in Figure~\ref{fig:jhhk}, tends to support the former possibility.
The spectral types of 2M~1101-7732 A and B are converted to 
effective temperatures with the temperature scale that was designed by
\citet{luh03b} to be compatible with the models of \citet{bar98} and 
\citet{cha00}. Bolometric luminosities are estimated by combining dereddened
$J$-band measurements, a distance of 168~pc \citep{whi97,wic98,ber99}, and
bolometric corrections described in \citet{luh99} and from \citet{rei01a}.
The combined uncertainties in $A_J$, $J$, and BC$_J$ ($\sigma\sim0.2$, 0.03,
0.1) correspond to errors of $\pm0.09$ in the relative values of
log~$L_{\rm bol}$.
When an uncertainty in the distance is included ($\sigma\sim10$~pc),
the total uncertainties are $\pm0.1$.
The extinctions, effective temperatures, and bolometric luminosities
for 2M~1101-7732 A and B are listed in Table~\ref{tab:data}.

\subsection{Ages and Masses}
\label{sec:mass}

\begin{figure}
\plotone{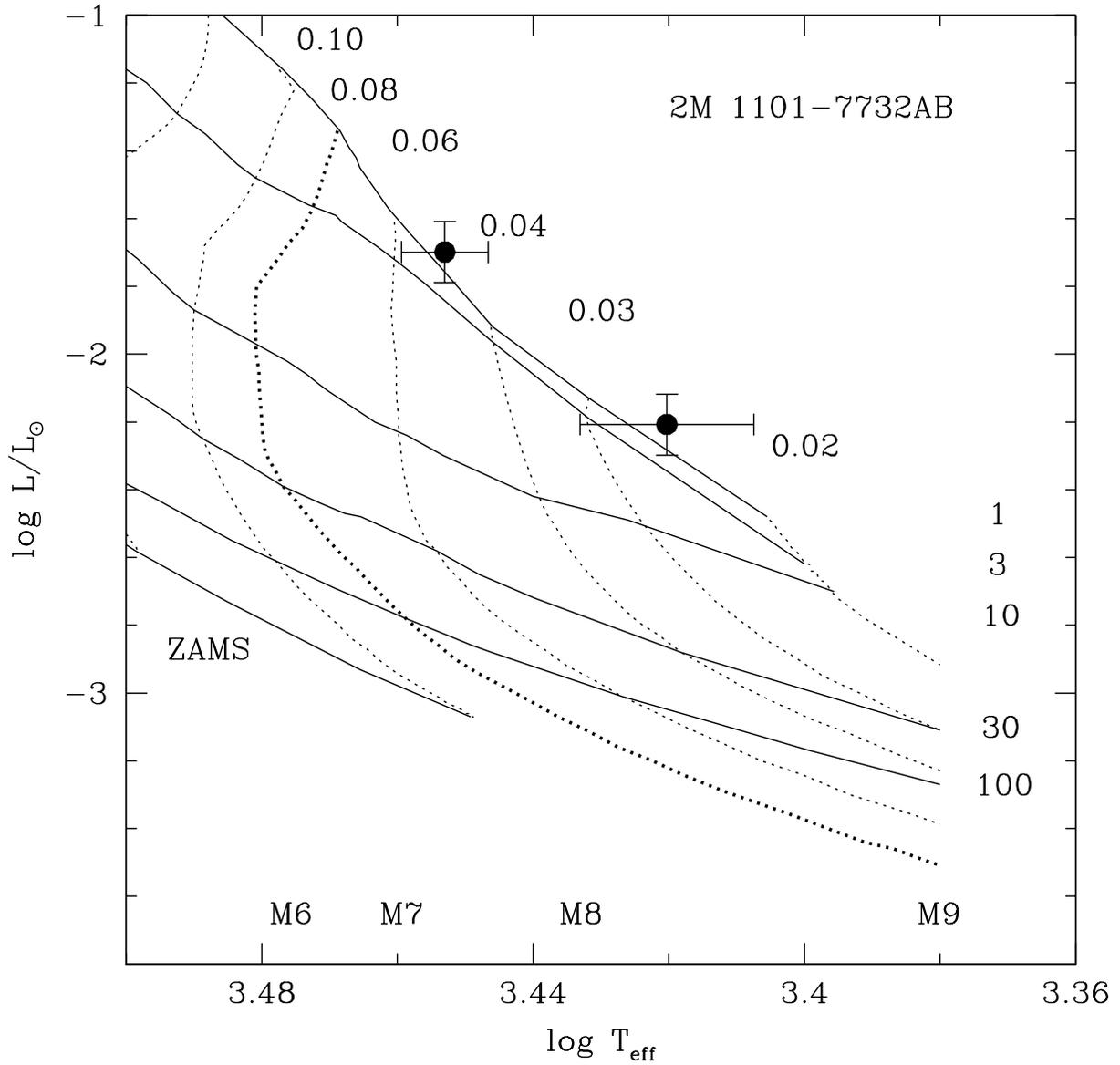}
\caption{
H-R diagram for 2M~1101-7732 A and B
shown with the theoretical evolutionary models of
\citet{bar98} ($M/M_\odot>0.1$) and \citet{cha00} ($M/M_\odot\leq0.1$),
where the mass tracks ({\it dotted lines}) and isochrones ({\it solid lines}) 
are labeled in units of $M_\odot$ and Myr, respectively. 
}
\label{fig:hr}
\end{figure}

The temperatures and luminosities derived in the previous section can be
interpreted in terms of ages and masses via theoretical evolutionary models.
For this analysis, I select the models of \citet{bar98} and \citet{cha00} 
because they provide the best agreement with observational constraints 
\citep{luh03b}.
The adopted temperature scale was designed to produce coevality for the 
components of the young quadruple GG~Tau and the same ages for the stellar 
and substellar members of Taurus and IC~348 with these 
evolutionary models \citep{luh03b}. 
As shown in the H-R diagram in Figure~\ref{fig:hr}, the components of 
2M~1101-7732AB exhibit nearly perfect coevality as well, providing strong 
support for the validity of this combination of temperature scale and models.
The coevality of 2M~1101-7732AB is additional evidence that these 
sources comprise a true binary system. 
The H-R diagram in Figure~\ref{fig:hr} implies masses of 
$0.05\pm0.01$ and $0.025\pm0.005$~$M_\odot$ for A and B, both of which are 
below the hydrogen burning mass limit.
The projected separation of $1\farcs44$ of these brown dwarfs corresponds
to 240~AU at the distance of Chamaeleon~I. Thus, 2M~1101-7732AB is the 
first known binary brown dwarf with a separation greater than $\sim20$~AU.

\section{Discussion}
\label{sec:dis}

Based on the absence of wide binary brown dwarfs in 
multiplicity surveys of the field and young clusters, 
\citet{bur03} suggested that wide systems do not form or are disrupted at 
ages of 1-10~Myr. The discovery of 2M~1101-7732AB now demonstrates that 
wide binary brown dwarfs indeed do exist. 
Because this system was found serendipitously and not through a
systematic companion search, it cannot be combined with previous multiplicity
surveys of brown dwarfs in star-forming regions (e.g., \citet{neu02})
to compute a frequency of wide binary brown dwarfs. Even if such an estimate
were possible, a single detection would not provide sufficient statistical
significance to determine if the frequency of wide systems differs
in star-forming regions and the field. However, the discovery of additional 
wide binary brown dwarfs in star-forming regions would indicate a clear
difference from the field, where none have been found with 2MASS among
the $\sim300$ known L and T dwarfs. 


The discovery of a wide binary brown dwarf has important implications for
understanding the origin of brown dwarfs. 
Previous constraints on the formation mechanism of substellar objects,
particularly the embryo ejection hypothesis, have been somewhat inconclusive. 
Detections of IR excesses and emission lines toward young objects near and 
below the hydrogen burning limit have indicated the
presence of circumstellar disks \citep{com98,com00,luh99,mue01,kle03},
accretion \citep{muz98,muz03,muz04,jay03,wb03,luh03a,bar03},
and outflows \citep{fer01,muz03,luh04b}, which have been taken as
evidence that brown dwarfs and stars might share a common formation history.
However, given that all of these signatures probe activity near the brown 
dwarfs, they do not exclude the ejection model, which predicts 
that only the outer portions of brown dwarf disks ($>20$~AU) are removed
during ejection \citep{bat03}. 
Meanwhile, the similarity between stars and brown dwarfs in their 
distributions of velocities \citep{jg01} and spatial positions 
\citep{bri02,luh03b,luh04b} contradicts the predictions of some ejection 
models \citep{rc01} but not others \citep{bat03}.
Finally, as pointed out in \S~\ref{sec:intro}, the binary fraction and maximum 
separation observed in previous surveys of field brown dwarfs are roughly 
consistent with the ejection hypothesis \citep{rc01,bat02}, 
but can be explained through other means as well \citep{bur03}.

In contrast to these previous results, 2M~1101-7732AB provides
arguably the most definitive insight to date into the formation of brown 
dwarfs. The existence of a 240~AU binary brown dwarf 
is clearly inconsistent with the ejection models, which predict a maximum 
separation of $\sim10$~AU for substellar binaries.
Any process that is capable of removing an embryo from
its parent core, envelope, and outer disk would easily disrupt a 
loosely bound pair of such embryos. As a result, ejection or other dynamical
processes could not have played a role in the formation of the brown dwarfs in
2M~1101-7732AB. 
Although some brown dwarfs may form through ejection, 
it is not an essential component of the birth of substellar objects.
Instead, it appears that brown dwarfs can arise from standard, unperturbed
cloud fragmentation.

\acknowledgements
I thank the staff of Las Campanas Observatory, particularly David Osip, 
for their assistance in these observations. I also thank Adam Burgasser for his
comments on the manuscript. 
This work was supported by grant NAG5-11627 from the NASA Long-Term Space 
Astrophysics program.

\begin{deluxetable}{lllllllllllll}
\tabletypesize{\scriptsize}
\tablewidth{0pt}
\tablecaption{Data for 2MASS J11011926-7732383AB\label{tab:data}}
\tablehead{
\colhead{} &
\colhead{} &
\colhead{} &
\colhead{} &
\colhead{$T_{\rm eff}$\tablenotemark{a}} &
\colhead{} &
\colhead{} &
\colhead{} &
\colhead{} &
\colhead{} &
\colhead{} &
\colhead{} &
\colhead{Mass} \\
\colhead{Component} &
\colhead{$\alpha$(J2000)} & 
\colhead{$\delta$(J2000)} &
\colhead{Spectral Type} &
\colhead{(K)} &
\colhead{$A_J$} & 
\colhead{$L_{\rm bol}$} & 
\colhead{$R-I$} & 
\colhead{$I$} &
\colhead{$J-H$} & 
\colhead{$H-K_s$} & 
\colhead{$K_s$} &
\colhead{($M_\odot$)}}
\startdata
A & 11 01 19.218 & -77 32 38.60 & M7.25$\pm0.25$ & 2838 & 0.45 & 0.020 & 2.57 & 17.21 & 0.84 & 0.60 & 11.97 & 0.05 \\
B & 11 01 19.438 & -77 32 37.36 & M8.25$\pm0.25$ & 2632 & 0 & 0.0062 & 2.60 & 18.28 & 0.82 & 0.64 & 12.81 & 0.025 \\
\enddata
\tablenotetext{a}{Temperature scale from \citet{luh03b}.}
\end{deluxetable}

\begin{thebibliography}{}

\bibitem[Baraffe et al.(1998)]{bar98}
Baraffe, I., Chabrier, G., Allard, F., \& Hauschildt, P. H. 1998, \aap, 337, 403

\bibitem[Barrado y Navascu\'es \& Mart{\'\i}n(2003)]{bar03}
Barrado y Navascu\'es, D., \& Mart{\'\i}n, E. L., 2003, \aj, 126, 2997

\bibitem[Basri(2000)]{bas00}
Basri, G. 2000, \araa, 38, 485

\bibitem[Bate et al.(2002)]{bat02}
Bate, M. R., Bonnell, I. A., \& Bromm, V. 2002, \mnras, 332, L65

\bibitem[Bate et al.(2003)]{bat03}
Bate, M. R., Bonnell, I. A., \& Bromm, V. 2003, \mnras, 339, 577

\bibitem[Bertout et al.(1999)]{ber99}
Bertout, C., Robichon, N., \& Arenou, F. 1999, \aap, 352, 574

\bibitem[Bessell \& Brett(1988)]{bb88}
Bessell, M. S., \& Brett, J. M. 1988, \pasp, 100, 1134

\bibitem[Boss(2001)]{bos01}
Boss, A. 2001, \apj, 551, L167

\bibitem[Bouy et al.(2003)]{bou03}
Bouy, H., Brandner, W., Mart{\'\i}n, E. L., Delfosse, X., Allard, F., \& Basri,
G. 2003, \aj, 126, 1526

\bibitem[Bouy et al.(2004)]{bou04}
Bouy, H., et al. 2004, \aap, in press

\bibitem[Brice\~no et al.(2002)]{bri02}
Brice\~{n}o, C., Luhman, K. L., Hartmann, L., Stauffer, J. R., \& Kirkpatrick,
J. D. 2002, \apj, 580, 317

\bibitem[Burgasser et al.(2003)]{bur03}
Burgasser, A. J., Kirkpatrick, J. D., Reid, I. N., Brown, M. E., Miskey, C. L.,
\& Gizis, J. E. 2003, \apj, 586, 512


\bibitem[Cambr\'esy et al.(1997)]{cam97}
Cambr\'esy, L., et al. 1997, \aap, 324, L5


\bibitem[Chabrier et al.(2000)]{cha00}
Chabrier, G., Baraffe, I. Allard, F., \& Hauschildt, P. H. 2000, \apj, 542, 464

\bibitem[Close et al.(2002a)]{clo02a}
Close, L. M., Potter, D., Brandner, W., Lloyd-Hart, M., Liebert, J.,
Burrows, A., \& Siegler, N. 2002a, \apj, 566, 1095

\bibitem[Close et al.(2003)]{clo03}
Close, L. M., Siegler, N., Freed, M., \& Biller, B. 2003, \apj, 587, 407

\bibitem[Close et al.(2002b)]{clo02b}
Close, L. M., Siegler, N., Potter, D., Brandner, W., \& Liebert, J. 2002b,
\apj, 567, L53


\bibitem[Comer\'{o}n et al.(2000)]{com00}
Comer\'{o}n, F., Neuh\"{a}user, R., \& Kaas, A. A. 2000, \aap, 359, 269

\bibitem[Comer\'{o}n et al.(1998)]{com98}
Comer\'{o}n, F., Rieke, G. H., Claes, P., Torra, J., \& Laureijs, R. J.
1998, \aap, 335, 522



\bibitem[Fern\'{a}ndez \& Comer\'{o}n(2001)]{fer01}
Fern\'{a}ndez, M., \& Comer\'{o}n, F. 2001, \aap, 380, 264

\bibitem[Freed et al.(2003)]{fre03}
Freed, M., Close, L. M., \& Siegler, N. 2003, \apj, 584, 453





\bibitem[Gizis et al.(2003)]{giz03}
Gizis, J. E., Reid, I. N., Knapp, G. R., Liebert, J., Kirkpatrick, J. D.,
Koerner, D. W., \& Burgasser, A. J. 2003, \aj, 125, 3302

\bibitem[Jayawardhana et al.(2003)]{jay03}
Jayawardhana, R., Mohanty, S., \& Basri, G. 2003, \apj, 592, 282

\bibitem[Joergens \& Guenther(2001)]{jg01}
Joergens, V. \& Guenther, E. 2001, \aap, 379, L9

\bibitem[Klein et al.(2003)]{kle03}
Klein, R., Apai, D., Pascucci, I., Henning, Th., Waters, L. B. F. M. 2003,
\apj, 593, L57

\bibitem[Koerner et al.(1999)]{koe99}
Koerner, D. W., Kirkpatrick, J. D., McElwain, M. W., \& Bonaventura, N. R.
1999, \apj, 526, L25



\bibitem[Kroupa \& Bouvier(2003)]{kb03b}
Kroupa, P., \& Bouvier, J. 2003, \mnras, 346, 369



\bibitem[Landolt(1992)]{lan92}
Landolt, A. U. 1992, \aj, 104, 340

\bibitem[Leggett(1992)]{leg92}
Leggett, S. K. 1992, \apjs, 82, 351

\bibitem[Luhman(1999)]{luh99}
Luhman, K. L. 1999, \apj, 525, 466


\bibitem[Luhman(2004a)]{luh04a}
Luhman, K. L. 2004a, \apj, 602, 816

\bibitem[Luhman(2004b)]{luh04b}
Luhman, K. L. 2004b, \apj, submitted



\bibitem[Luhman et al.(2003a)]{luh03a}
Luhman, K. L., Brice\~{n}o, C., Stauffer, J. R., Hartmann, L.,
Barrado y Navascu\'{e}s, D., \& Nelson, C. 2003a, \apj, 590, 348

\bibitem[Luhman et al.(2003b)]{luh03b}
Luhman, K. L., Stauffer, J. R., Muench, A. A., Rieke, G. H., Lada, E. A.,
Bouvier, J., \& Lada, C. J. 2003b, \apj, 593, 1093



\bibitem[Mart{\'\i}n et al.(1998)]{mar98}
Mart{\'\i}n, E. L., et al. 1998, \apj, 509, L113

\bibitem[Mart{\'\i}n et al.(2000)]{mar00}
Mart{\'\i}n, E. L., et al. 2000, \apj, 543, 299

\bibitem[Mart{\'\i}n et al.(2003)]{mar03}
Mart{\'\i}n, E. L., Barrado y Navascu\'{e}s, D., Baraffe, I., Bouy, H.,
\& Dahm, S. 2003, \apj, 594, 525

\bibitem[Mart{\'\i}n et al.(1999)]{mar99}
Mart{\'\i}n, E. L., Brandner, W., \& Basri, G. 1999, Science, 283, 1718

\bibitem[Mart{\'\i}n et al.(1996)]{mar96}
Mart{\'\i}n, E. L., Rebolo, R., \& Zapatero Osorio, M. R. 1996, \apj, 469, 706

\bibitem[Muench et al.(2001)]{mue01}
Muench, A. A., Alves, J., Lada, C. J., \& Lada, E. A. 2001, \apj, 558, L51

\bibitem[Muzerolle et al.(1998)]{muz98}
Muzerolle, J., Hartmann, L., \& Calvet, N. 1998, \aj, 116, 455

\bibitem[Muzerolle et al.(2003)]{muz03}
Muzerolle, J., Hillenbrand, L., Calvet, N., Brice\~{n}o, C., \& Hartmann, L.
2003, \apj, 592, 266

\bibitem[Muzerolle et al.(2004)]{muz04}
Muzerolle, J., et al. 2004, in preparation


\bibitem[Neuh\"{a}user et al.(2002)]{neu02}
Neuh\"{a}user, R., Guenther, E., Mugrauer, M., Ott, T., \& Eckart, A. 2002,
\aap, 395, 877




\bibitem[Reid et al.(2001a)]{rei01a}
Reid, I. N., Burgasser, A. J., Cruz, K. L., Kirkpatrick, J. D., \& Gizis, J. E.
2001a, \aj, 121, 1710

\bibitem[Reid et al.(2001b)]{rei01b}
Reid, I. N., Gizis, J. E., Kirkpatrick, J. D., \& Koerner, D. W.  2001b, \aj,
121, 489



\bibitem[Reipurth \& Clarke(2001)]{rc01}
Reipurth, B.~\& Clarke, C.\ 2001, \aj, 122, 432


\bibitem[White \& Basri(2003)]{wb03}
White, R. J., \& Basri, G. 2003, \apj, 582, 1109


\bibitem[Whittet et al.(1997)]{whi97}
Whittet, D. C. B., Prusti, T., Franco, G. A. P., Gerakines, P. A.,
Kilkenny, D., Larson, K. A., \& Wesselius, P. R. 1997, \aap, 327, 1194

\bibitem[Wichmann et al.(1998)]{wic98}
Wichmann, R., Bastian, U., Krautter, J., Jankovics, I., \& Ruci\'nski, S. M.
1998, \mnras, 301, L39

\end{thebibliography}
\end{document}